\documentclass[%
 reprint,
superscriptaddress,
 amsmath,amssymb,
 aps,
prl,
]{revtex4-1}

\usepackage{graphicx}
\usepackage[dvipsnames]{xcolor}
\usepackage{amssymb}
\usepackage{amsmath}
\usepackage{amsbsy}
\usepackage{color}

\usepackage{bm}

\usepackage{siunitx}

\begin{document}

\title{Geometric Learning of Knot Topology}

\author{Joseph Lahoud Sleiman}
\affiliation{School of Physics and Astronomy, University of Edinburgh}
\thanks{joint first author}

\author{Filippo Conforto}
\affiliation{School of Physics and Astronomy, University of Edinburgh}
\thanks{joint first author}

\author{Yair Augusto Gutierrez Fosado}
\affiliation{School of Physics and Astronomy, University of Edinburgh}

\author{Davide Michieletto}
\affiliation{School of Physics and Astronomy, University of Edinburgh}
\affiliation{MRC Human Genetics Unit, Institute of Genetics and Cancer, University of Edinburgh}
\thanks{corresponding author}

\begin{abstract}
Knots are deeply entangled with every branch of science. One of the biggest open challenges in knot theory is to formalise a knot invariant that can unambiguously and efficiently distinguish any two knotted curves. Additionally, the conjecture that the geometrical embedding of a curve encodes information on its underlying topology is, albeit physically intuitive, far from proven. Here we attempt to tackle both these outstanding challenges by proposing a neural network (NN) approach that takes as input a geometric representation of a knotted curve and tries to make predictions of the curve's topology. Intriguingly, we discover that NNs trained with a so-called geometrical ``local writhe'' representation of a knot can distinguish curves that share one or many topological invariants and knot polynomials, such as mutant and composite knots, and can thus classify knotted curves more precisely than some knot polynomials. Additionally, we also show that our approach can be scaled up to classify all prime knots up to 10-crossings with more than 95\% accuracy. Finally, we show that our NNs can also be trained to solve knot localisation problems on open and closed curves. Our main discovery is that the pattern of ``local writhe'' is a potentially unique geometric signature of the underlying topology of a curve. We hope that our results will suggest new methods for quantifying generic entanglements in soft matter and even inform new topological invariants.
\end{abstract}

\maketitle

\section{Introduction}
Knots are fascinating objects that have captured the attention of humans for centuries. From Incas' knotted Quipus~\cite{Adams1994}, and Lord Kelvin's theory of elements as knotted ether~\cite{KelvinAtoms}, to sailors and climbers whose lives often rely on the strength of knotted rope, knots are deeply intertwined with history and art and often carry mystical meaning. The human obsession with knots brought Peter Guthrie Tait to compile the first knot tabulation of up to 10 crossings by hand~\cite{Tait1885}; currently, more than one million unique knots up to 16 crossings have been tabulated using computer programs~\cite{Hoste1998}.

To rigorously prove that the early tabulated knots did not contain duplicates, so-called topological invariants and knot polynomials were developed, the first of which was the Alexander polynomial~\cite{Alexander1928, Kauffman1983, Adams1994}, followed more recently by the Jones and HOMFLY polynomials~\cite{Hoste1986,Adams1994}. Knot polynomials are mathematical constructs that can be computed on knot diagrams and are invariant under smooth deformations of the curve, i.e. deformations that preserve the curve topology. 
However, there are knots that share many topological invariants and cannot even be distinguished by knot polynomials. Famously, the 11-crossing Conway knot has the same Alexander polynomial as the unknot and shares the same Jones polynomial of its mutant, the Kinoshita–Terasaka (KT) knot~\cite{Adams1994}. More generally, all mutants of a knot have the same HOMFLY polynomials and the same hyperbolic volume~\cite{Adams1994}, while some composite knots share the same homeomorphic complements~\cite{Thurston1997,SnapPy,Gordon1989}.

\begin{figure*}[t!]
  \centering
  \includegraphics[width=0.95\textwidth]{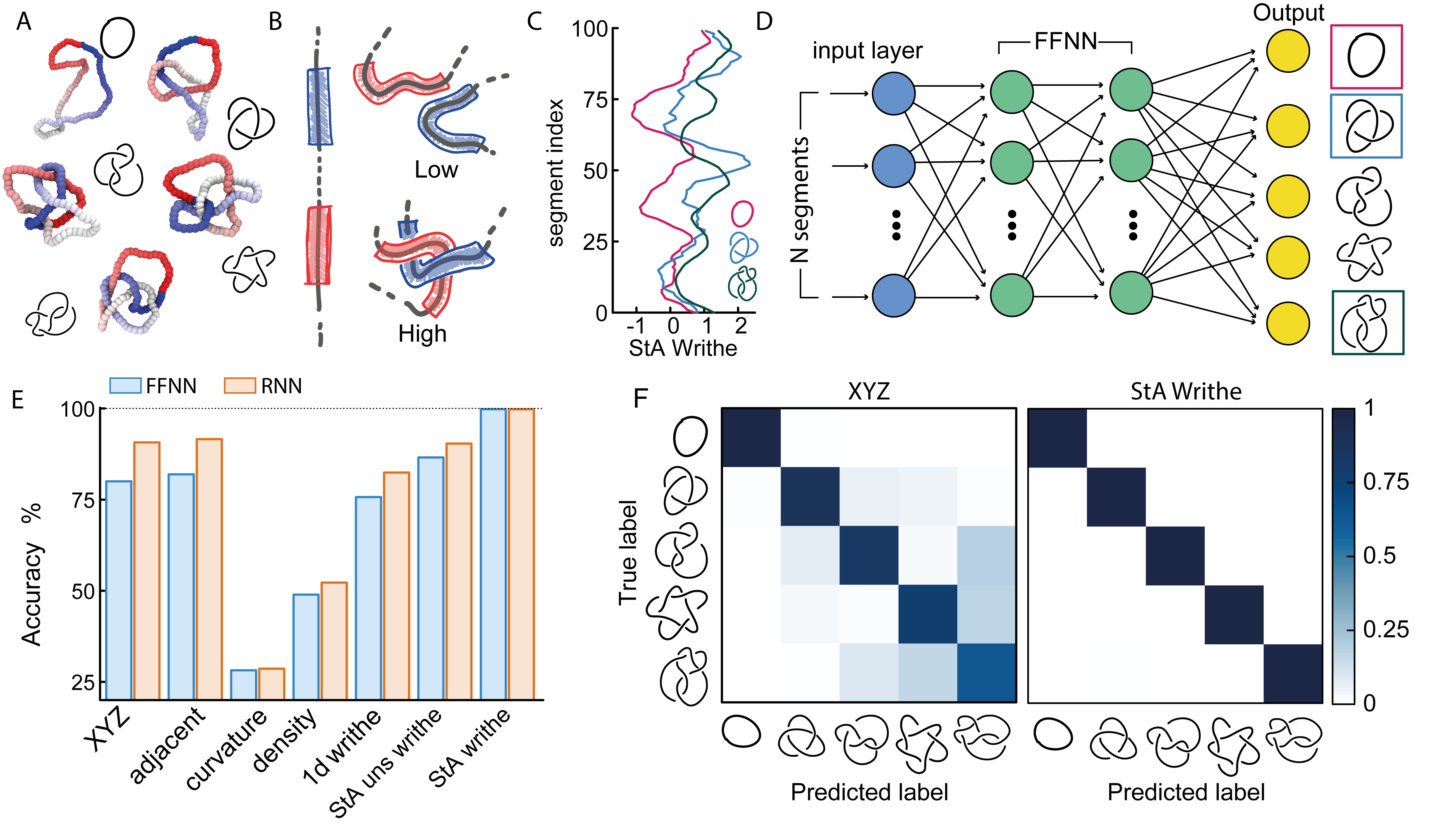}
\caption{\textbf{A} Examples of equilibrium knotted polymer conformations colour coded to show the knot contour (red, white, and blue). In this figure we consider the 5 simplest knots: $0_1$, $3_1$, $4_1$, $5_1$ and $5_2$. \textbf{B} A graphical representation of StS writhe $\omega_{StS}(x,y)$ showing a situation of low and high writhe between two segments. \textbf{C} Examples of patterns for $\omega_{StA}(x)$ for three different knots. \textbf{D} A graphical representation of the (feed-forward) network. The input layer contains $N$ (or $3N$) neurons corresponding to the size of the input feature representation, and the output layer yields a probability for each knot class. \textbf{E} Accuracy score, tested on unseen polymer conformations for different input features. The StA writhe classifies the 5-simplest knots with 99.9\% accuracy irrespectively of the network architecture. \textbf{F} Confusion matrices obtained by training the network with XYZ and StA writhe input features.
}
\label{fig:panel0}
\end{figure*}

Alongside the development of topological invariants, several attempts were made to identify a relationship between a specific \textit{geometric} embedding of a knot and its underlying topology~\cite{stasiak1998ideal}. We note that this relationship is different from the one sought between so-called geometric and algebraic invariants~\cite{Jejjala2019, Davies2021}, e.g. between the hyperbolic volume of a knot and its Jones polynomial~\cite{Murakami2002}. Perhaps one of the most rigorous results in this direction is the F\'{a}ry-Milnor theorem, stating that the total absolute curvature of non-trivially knotted curves must be greater than $4 \pi$~\cite{Milnor1950}. Unfortunately, this result only imposes a weak constraint on the topology of the underlying curve, as an unknot can itself have large total curvature due to, for example, deformations of its contour. 
In parallel, a large body of work on so-called ``ideal knots'' was done with the aim of finding geometric features that could reflect the underlying knot topology. One impressive result in this context is that different DNA knots display a spatial separation when run on a gel electrophoresis that is linearly proportional to the so-called average crossing number~\cite{Stas,Michieletto2015gelep}; this result entails that there is an intimate relationship between the physical shapes assumed by knots and their underlying topology. Another result that inspired our work is that the total so-called ``writhe'' (see below) of an ideal knot is the same (up to a constant that is only a function of the curve length) as that of a non-ideal, thermally agitated curve with the same topology~\cite{Katritch1997}. Though this suggests that ``writhe'' may be a good measure that is invariant under thermal fluctuations, there is no one-to-one relationship between the global writhe of a knot and its underlying topology; for instance, the global writhe of the $4_1$ knot is 0, the same as the unknot~\cite{stasiak1998ideal}.

Thus, the problem of determining a curve topology only based on the geometric information of its segments (without using any projection and algebraic invariant) is an open challenge in knot theory that has ramification in many fields, for instance polymer physics, biophysics and fluid dynamics. In this paper, we propose to address this open challenge by using the power of artificial intelligence, and in particular deep learning, at recognising and classifying patterns in certain knot geometric features. Our main discovery is that by using a quantity we dub ``local writhe'', even simple machine learning (ML) algorithms can identify the topology of knotted curves undergoing thermal fluctuations with more than 99\% accuracy. We argue that this is an example of geometric learning as the only quantity we pass to the ML algorithm is a quantity that can be computed from the Cartesian positions of a curve's segments, without the need of computing algebraic invariants such as Alexander or Jones polynomials. Our method can even distinguish 11-crossing knots that are otherwise impossible to distinguish using standard invariants (the Conway and KT knots). 
Finally, we show how this algorithm can be scaled to classify all 250 prime knots up to 10 crossings with 95\% accuracy and that can even be employed to solve knot localisation problems. Overall, we argue that local writhe is an excellent feature -- based purely on the 3D position of a curve segments -- that displays patterns that can be easily identified by ML algorithms. We hope that out results will be applied to other classification problems such as threading~\cite{Michieletto2016pnas,Stano2022} and entanglements~\cite{Everaers2004,smrek2021topological} and also prompt knot theorists to employ local writhe to define new geometric knot invariants.

\section{Results}

Two recent papers by Vandans~\cite{Vandans2020} and Braghetto~\cite{Braghetto2022} have shown that machine learning is a promising tool to solve knot classification problems. They mostly considered Cartesian position of the monomers or adjacent monomer distances and dihedrals to classify the 5 simplest knots. In this work, we set out to test the use of a different type of geometric features that our group recently considered to identify essential crossings of a knot and plectoneme-like double folding of ring polymers~\cite{Michieletto2016tree,Sleiman2022}. More specifically, we focused on a generalisation of the Gauss linking integral applied to a single closed curve, often associated with its writhe~\cite{Dennis2005} and average crossing number~\cite{Stasiak1996,Michieletto2016tree}. This choice is inspired by the intuition that writhe captures the geometrical entanglement of a curve with itself, and we thus define a generalised local segment-to-segment (StS) writhe as  
\begin{equation}
\omega_{StS}(x,y) =  \dfrac{(\bm{t}(x) \times \bm{t}(y)) \cdot (\bm{r}(x) - \bm{r}(y))}{|\bm{r}(x) - \bm{r}(y)|^3}  \, , 
 \label{eq:stswrithe}
\end{equation}
where $\bm{r}(x)$ and $\bm{t}(x)$ are the 3D position of, and the tangent at, segment $x$, respectively. Intuitively, Eq.~\eqref{eq:stswrithe} captures the magnitude and the chirality of the entanglement between segment $x$ and segment $y$ (Fig.~\ref{fig:panel0}A-B). The quantity $\omega_{StA}(x) = \oint_\gamma \omega_{StS}(x,y) dy$ is the local segment-to-all (StA) writhe and characterises how geometrically entangled segment $x$ is with respect to the whole closed curve $\gamma$. In practice, the calculation of StS and StA writhe are conducted on discrete segments, taking a finite ``window'' with length $l_w = 10$ $\sigma$ to smooth out short length fluctuations (see SI for details).

The StA writhe, $\omega_{StA}(x)$, is a 1D geometrical representation of a knot that we hypothesise may display some patterns that are topology-dependent (Fig.~\ref{fig:panel0}A-C). Since complex pattern recognition is a task that naturally lends itself to being addressed using a machine learning approach, we thus asked ourselves if a neural network (NN) trained to recognise patterns within $\omega_{StA}(x)$ was able to solve ambiguous knot classification problems. To do this, we built feed forward and recurrent (long-short term memory, LSTM) neural networks (FFNN and RNN, respectively) and trained them using $10^5$ statistically uncorrelated and pre-labelled conformations for each knot. To generate these conformations, we initialised a bead-spring polymer with known topology, $N=100$ beads, and persistence length $l_p=10 \sigma$ (other lengths and $l_p$ are reported in the SI) using KnotPlot (\url{knotplot.com}) and subsequently evolved the polymer configurations in LAMMPS~\cite{Plimpton1995} via Langevin dynamics in an implicit solvent and fixed temperature and using a Kremer-Grest model~\cite{Kremer1990} to preserve polymer topology (see Methods and SI for more details). The code to generate these conformations are open access at \url{https://git.ecdf.ed.ac.uk/taplab/mlknotsproject}. We confirmed that the topology was conserved either by computing their Alexander determinant via KymoKnot (\url{kymoknot.sissa.it})~\cite{Tubiana2018} or, when ambiguous, visually. 

\begin{figure*}[t!]
    \centering
    \includegraphics[width=0.95\textwidth]{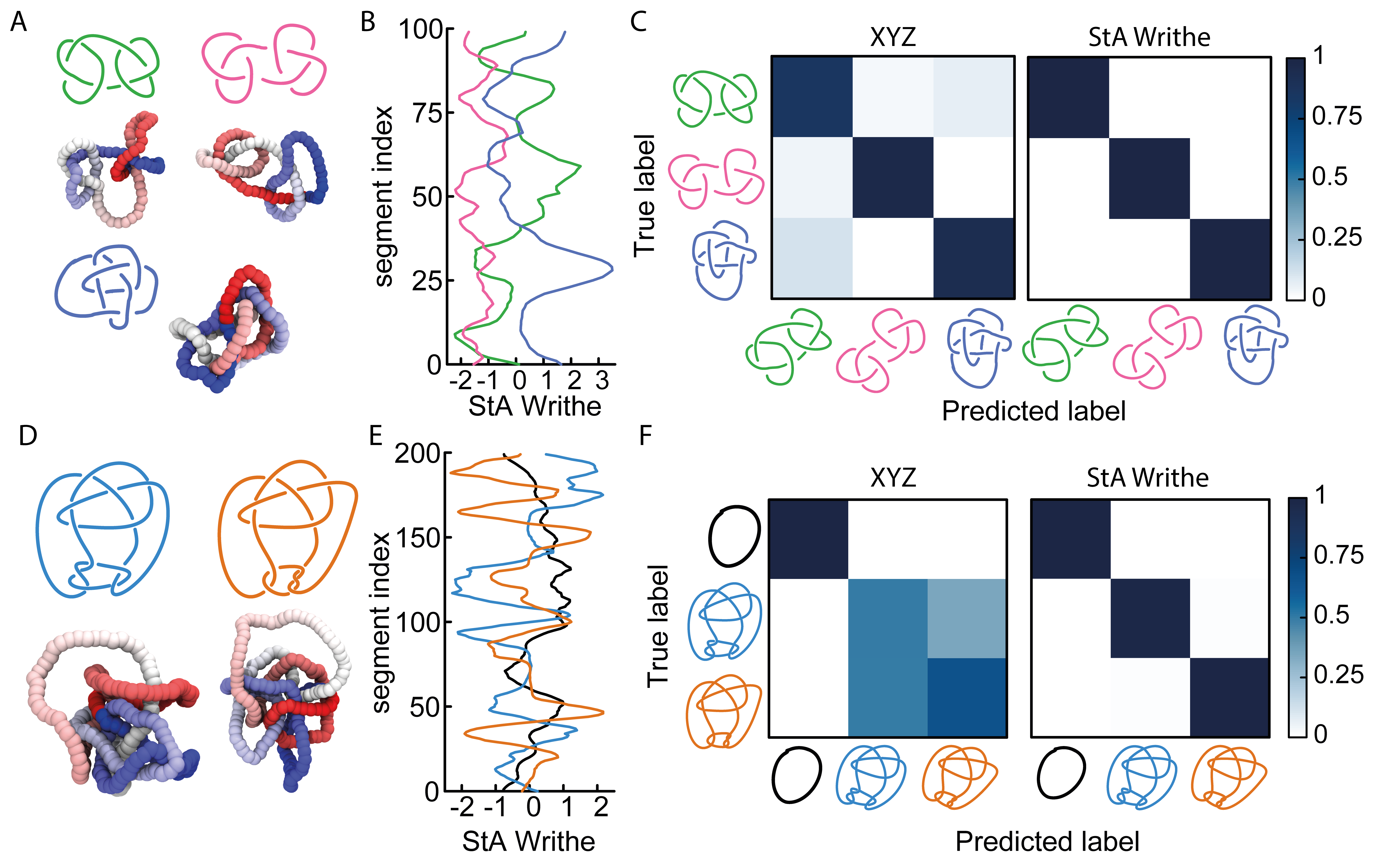}
    \caption{ \textbf{A} Snapshots of three knots with identical Alexander polynomial: square ($3^l_1\# 3^r_1$), granny ($3^l_1 \# 3^l_1$) and $8_{20}$ knots. \textbf{B} Examples of StA writhe patterns from the three knots. \textbf{C} Confusion matrices obtained from a 3-class classification problem, training a FFNN with XYZ (91.7\% accuracy) or StA writhe (99.9\% accuracy) features. \textbf{D} Snapshots of Conway (blue) and KT (orange) knots. \textbf{E} Examples of StA writhe patterns, including the one from the unknot (black). \textbf{F} Confusion matrices obtained from a 3-class classification problem training a FFNN with XYZ (67\% accuracy) and StA writhe (99.6\% accuracy). 
    }
    \label{fig:panel1}
\end{figure*}

\begin{figure*}[t!]
    \centering
    \includegraphics[width=0.95\textwidth]{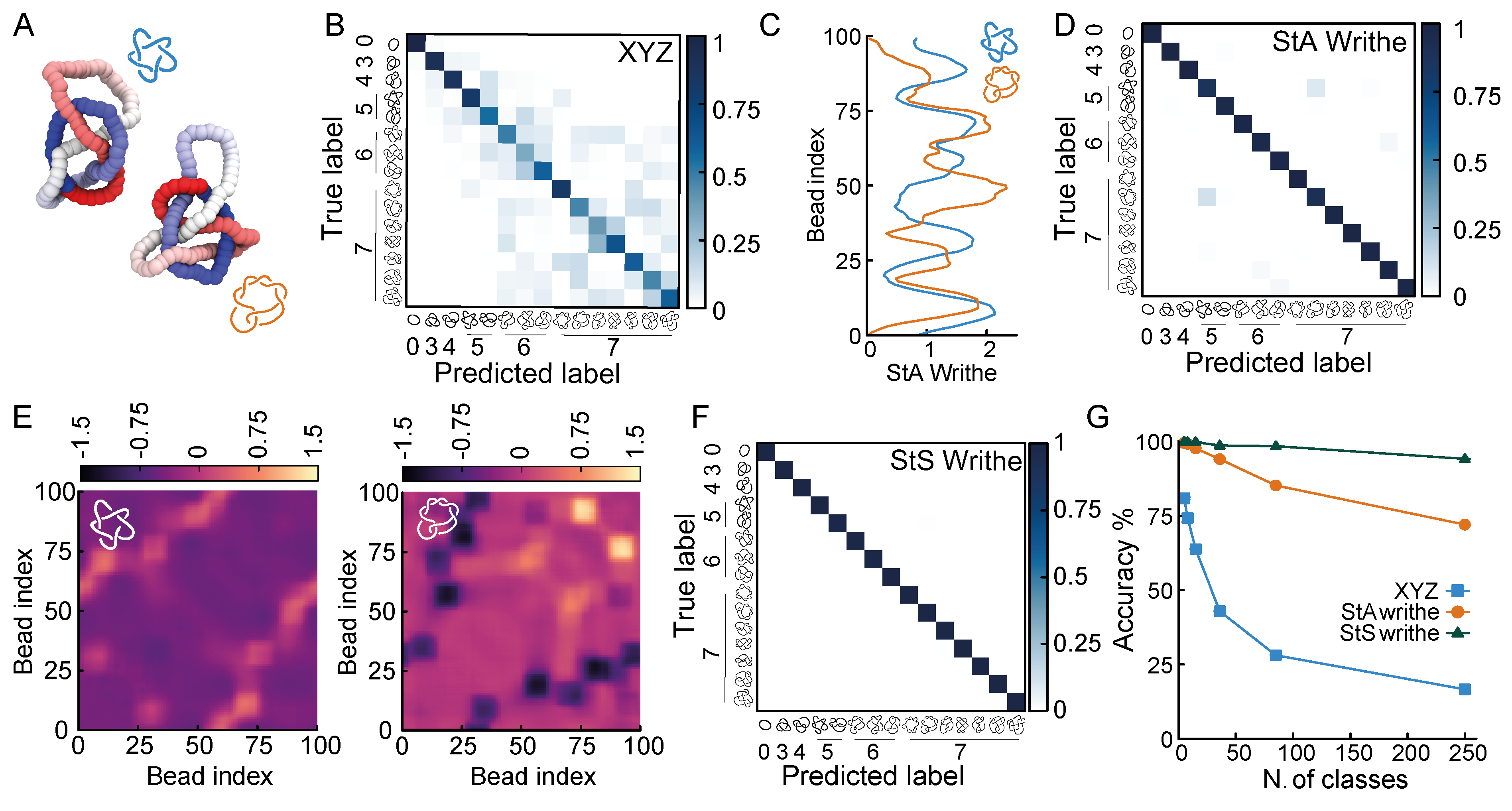}
    \caption{\textbf{A} Two example conformations of $5_1$ (blue) and $7_2$ (orange) knots. \textbf{B} The XYZ-trained NN on a 15-class classification problem yields 63.8\% accuracy and a rather non-diagonal confusion matrix. \textbf{C} Examples of $\omega_{StA}(x)$ curves for the two knots, displaying a degree of similarity between the pattern of maxima and minima. \textbf{D} The $\omega_{StA}(x)$-trained NN achieves 98\% accuracy and the confusion matrix shows that $5_1$ and $7_2$ are the knots that are most confused with each other. \textbf{E} Examples of $\omega_{StS}(x,y)$ geometric feature for the two knots corresponding to the $\omega_{StA}(x)$ profiles shown in \textbf{C}. \textbf{F} Confusion matrix for a StS-trained FFNN, achieving 99.8\% accuracy. \textbf{G} Accuracy as a function of number of knot classes to be distinguished, up to 10-crossing (250) prime knots. }
    \label{fig:panel2}
\end{figure*}

The NNs were built with an input layer that was determined according to the input representation being studied, e.g., the Cartesian (XYZ) coordinate representation used 3 neurons (one for each dimension) per polymer bead. Other local input features, such as StA writhe, used one neuron per bead, while the StS writhe feature requires $N \times N$ input neurons. The optimal number of hidden layers, hidden units, learning rate and batch size were determined via an automated hyperparameter tuning method conducted on the Cartesian representation (\textit{KerasTuner}~\cite{omalley2019kerastuner}). Unless otherwise stated, our NNs contained 4 hidden layers, with around $4 \cdot 10^5$ trainable parameters. The output layer consisted of $C$ output neurons, corresponding to the $C$ knot types being classified, each implemented with a softmax activation function in order to return the probability that a given input is a certain knot type. We took the sparse categorical cross-entropy as the loss function, as the most appropriate for individual class probabilities and integer target labels, i.e. our knot types (Fig.~\ref{fig:panel0}D).

\subsection{StA writhe yields NNs that are more accurate to classify the simplest 5 knots than Cartesian features.}

We first tackle a 5-knot classification problem with the 5 simplest knots, which can be satisfactorily solved using NNs trained on center-of-mass-corrected Cartesian coordinates (XYZ) or adjacent bead input features~\cite{Vandans2020,Braghetto2022}. In line with these previous works, we find that our NNs can accurately predict the topology of unseen conformations (80.1\% accuracy with a FFNN and 86\% accuracy with a recurrent NN architecture, Fig.~\ref{fig:panel0}E). These values are lower than the ones reported in Ref.~\cite{Vandans2020} because we use a smaller training dataset and smaller NNs. We then trained the same NNs using a range of other geometric features, such as local curvature, density and 1D writhe~\cite{Sleiman2022} (see SI for details), and found that most of them performed more poorly, or at best equally, with respect to the XYZ representation (Fig.~\ref{fig:panel0}E). A similar outcome was also obtained in Ref.~\cite{Braghetto2022}. In stark contrast, models trained using $\omega_{StA}(x)$ outperformed all other models and are found to achieve 99.9\% accuracy, irrespective of the FFNN or RNN architectures (we also tested random forest algorithms, see SI). Additionally, the networks reached the early stopping criterion in about 50\% fewer epochs or less, compared to those trained using the XYZ representation (see SI). When plotted as a confusion matrix, the results clearly indicate that the XYZ input feature struggles to classify knots with a similar number of crossings, e.g. the $5_1$ and $5_2$ knots. In contrast, our local 3D writhe (StA) feature generated a near-perfect confusion matrix (Fig.~\ref{fig:panel0}F).

We found that these results are generally robust for different choices of dataset splitting, persistence length ($l_p = 1-10$ $\sigma$), window length chosen to perform the StA calculation, and length of the chains (see SI). Nevertheless, they do display a significant reduction in accuracy when tested on knots generated using a different method (for instance freely jointed chains), and also when the window length for the StA writhe calculation is comparable to the full contour of the chain. In this case, the StA writhe is constant and equals the global writhe of the knot, which is not unique for different knots~\cite{stasiak1998ideal}. This is also in agreement with a principal component analysis (PCA, see SI), where we see that different knots are clearly separable in the reduced 2D PCA space, yet the $0_1$ and $4_1$ cluster together due to the fact that they share the global writhe (zero), which is related to the mean value of $\omega_{StA}(x)$ along the contour. 

\subsection{NNs trained with StA writhe can distinguish knots with identical knot polynomials}

Given that our NNs distinguished knots with same minimal number of crossings, i.e., the 5-crossings knots, we ask ourselves if they could also solve more complicated problems where knots shared algebraic knot polynomials. To this end, we first considered three knots with identical Alexander polynomial: the square, granny, and $8_{20}$ knots (see Fig.~\ref{fig:panel1}A). The first two knots are 6-crossings knots consisting of trefoil composites with different chirality (hence they are homeomorphic knot complements), whereas the latter is an 8-crossings knot. Once again, we trained our FFNN using the $\omega_{StA}(x)$ profiles (Fig.~\ref{fig:panel1}B) and obtained a striking accuracy of 99.98\%, compared with 91.8\% obtained by training with COM-shifted XYZ coordinates (Fig.~\ref{fig:panel1}C). 

We then asked ourselves if our method could also perform just as well in situations where the knots shared multiple knot polynomials. As mentioned above, mutant knots share the same hyperbolic volume and several knot polynomials, including HOMFLY. We therefore performed simulations of the Conway (K11n34) knot and one of its mutants, the Kinoshita–Terasaka (KT, K11n42) knot. These 11-crossings knots have a number of identical knot invariants as they share the same Jones, Alexander, and Conway polynomials~\cite{Adams1994}. Intriguingly, the latter two are also shared with the unknot. Thus, we generated $10^5$ statistically uncorrelated conformations of $N=200$ beads long polymers with the Conway, KT, and unknot topologies (Fig.~\ref{fig:panel1}D), and trained our FFNN to classify them either using a COM-subtracted XYZ or $\omega_{StA}(x)$ (Fig.~\ref{fig:panel1}E) representations. When tested on unseen conformations, we found that the XYZ-trained NN could not distinguish the Conway and KT knots, but both are accurately distinguished from the unknot (Fig.~\ref{fig:panel1}F). In marked contrast, we discovered that the StA-trained NN perfectly disentangles the three knots with 99.6\% accuracy (Fig.~\ref{fig:panel1}F). We therefore conclude that the StA-trained NN has the ability to convert StA patterns into a topological knot classification, even for knots sharing multiple knot polynomials, such as mutants and composites. In turn, we argue that the StA writhe is a geometric quantity computed on a particular 3D embedding of a curve that carries high-density information about its underlying topology. Importantly, we stress that to classify these knots the network does not compute any knot polynomial, as other standard software do.

Somewhat unsatisfactorily, we cannot fully pinpoint why StA writhe is so powerful at identifying different topologies. We hypothesise that the 1D patterns generated by StA writhe, and specifically the sequence, sign and amplitudes assumed by consecutive maxima and minima, contain information on the relative orientation and severity of consecutive entanglements. As mentioned above, the average value of $\omega_{StA}(x)$ is related to the global writhe of the knot, which itself contains non-unambiguous information on its topology. Thus, we argue that the NNs can extract additional information from the full $\omega_{StA}(x)$ patterns that is related to the chirality of individual entanglements and render the information unique. This hypothesis is also supported by the fact that considering \textit{unsigned} StA writhe (which cannot distinguish chirality) yields in general a lower accuracy (see Fig.~\ref{fig:panel0}E). We thus hypothesise that the information encoded in the pattern of the StA writhe may be related to the underlying knot's Dowker code. These hypotheses will be tested in more detail in future works.

\subsection{StS writhe outperforms StA writhe on knots with more than 7 crossings}

To understand to what extent StA-trained NNs can be used to classify knotted curves, we trained our NNs on increasingly more complex classification problems, and generated conformations of all prime knots up to 10-crossings. Among these 250 prime knots, there are over 30 that share the same Alexander polynomial (see SI for a table), making them challenging to classify using standard tools (for instance Kymoknot). We first noticed that XYZ-trained NNs rapidly declined in accuracy when we included knots with 6 or more crossings (Fig.~\ref{fig:panel2}A-B). In contrast, the confusion matrices from StA-trained NNs retained relatively high accuracies. However, we noticed that the knots $5_1$ and $7_2$ created some confusion even in the StA-trained NNs, causing a drop in accuracy to 98\% (Fig.~\ref{fig:panel2}C-D). We argue that this was due to the fact that $\omega_{StA}(x)$ displayed similar patterns between the two knot types. For instance, we show two knots that yield particularly similar $\omega_{StA}(x)$ patterns in Fig.~\ref{fig:panel2}C. Thus, to further distinguish these (and potentially other knots with similar $\omega_{StA}(x)$ curves) we decided to consider our original proposition of using the local StS writhe (Eq.~\eqref{eq:stswrithe}); two examples of $\omega_{StS}(x,y)$ maps are reported in Fig.~\ref{fig:panel2}E, for the same $5_1$ and $7_2$ knots configurations used to compute $\omega_{StA}(x)$ in Fig.~\ref{fig:panel2}C. Interestingly, the $\omega_{StS}(x,y)$ maps appear very different, despite generating very similar StA curves when integrated along $y$ and around the polymer contour. This is because a given segment $x$ may itself have a certain sequence of negative and positive entanglements with other segments $y$. Once integrated along the contour in the $y$ direction, different sequences may lead to similar overall values. 
Motivated by this, we trained our FFNNs using the StS writhe representation of the knots, and discovered we could restore a very high (99.8\%) accuracy for the case of a database containing all knots up to 7-crossings (Fig.~\ref{fig:panel2}F). More specifically, the confusion between $5_1$ and $7_2$ knots is now solved thanks to the StS writhe. Ultimately, the StS-trained NNs produced the most accurate models, achieving 95\% for a 250-class classification task including all knots up to 10 crossings. In comparison, the XYZ-trained and StA-trained NNs achieved 17\% and 72\% on the same problem, respectively (Fig.~\ref{fig:panel2}G).

Based on these results, we argue that the StS writhe is therefore the most scalable and precise geometric feature to employ for knot classification problems. Most importantly, we stress that the impressive accuracy for a 250-class problem was achieved with a simple feed forward NN with 4 layers (around 400k parameters). A natural extension going forward will be to employ more complex architectures and in particular convolutional NNs to classify the 2D StS writhe maps. 

\begin{figure*}[t!]
    \centering
    \includegraphics[width=0.95\textwidth]{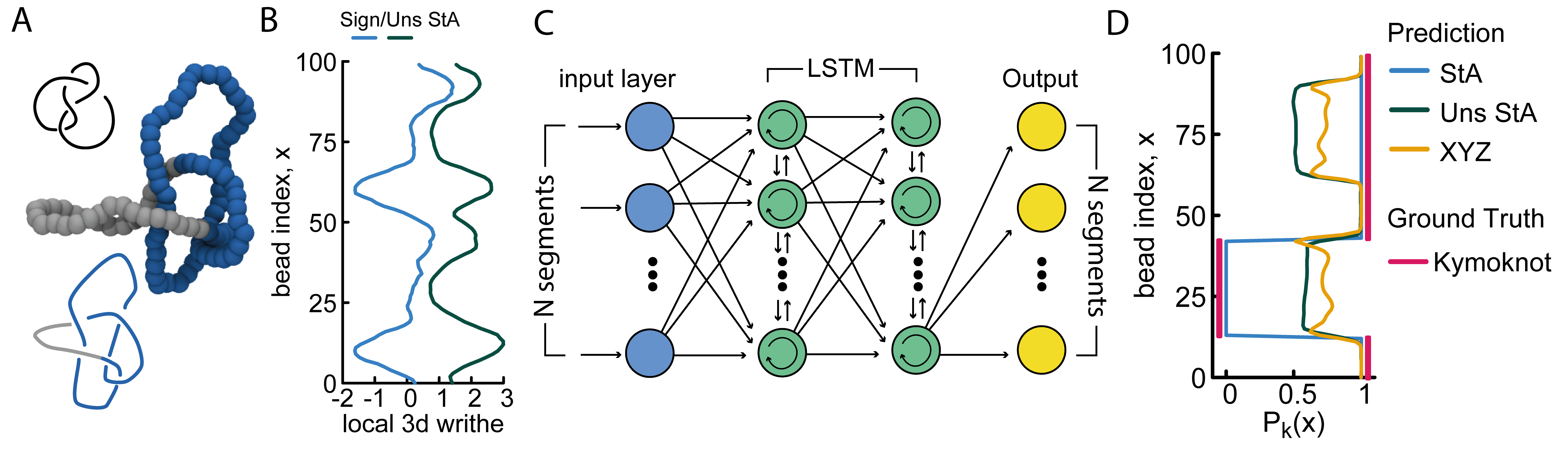}
    \caption{\textbf{A} Example of a $4_1$ knot where the knotted core is localised within $\sim$ 80\% of the contour.\textbf{B} Signed and unsigned StA writhe profiles for the conformation shown in \textbf{A}. \textbf{C} Sketch of a LSTM (recurrent) NN, encoding the sequential information of the segments. \textbf{D} Profile of the knot probability $P_k(x)$ as a function of bead index $x$, as predicted by the RNN with different geometric features. The ground truth was generated using KymoKnot.}
    \label{fig:loc_acc}
\end{figure*}

\subsection{StA-trained NNs can also solve knot localisation problems.}

In the final part of this paper we turn our attention to the knot localisation problem, i.e. determining the shortest knotted arc along the polymer contour. This task is challenging and particularly important for open curves, such as linear polymers, DNA, and proteins~\cite{Dabrowski-Tumanski2016,Giulini2019,Klotz2018,Caraglio2019,Caraglio2020,Soh2019}, which may contain entanglements and knots. In this context, identifying the shortest portion of a polymer that is knotted is akin to being able to identify entanglements in chain melts. 

We first tackled this problem using the same FFNN architecture as in the knot classification task, but the accuracies generated were very low. We hypothesised that this was due to the fact that FFNN do not preserve the sequential information along the polymer. For this reason, we consider a long-short term memory (LSTM), also known as recurrent NN (RNN). More specifically, we employed a sequence-to-sequence LSTM, with an output layer corresponding to a binary sequence of $N=100$ neurons, equivalent in dimension to the length of the input polymer. Each output neuron is passed through a sigmoid function, which converts the output into a probability between 0 and 1 representing the likelihood that a given monomer is within the knotted segment of the polymer conformation. The true output labels were generated using KymoKnot~\cite{Tubiana2018}, which employs a minimally-interfering closure algorithm followed by a standard Alexander determinant calculation to identify the start and end monomers of the knot. This data was then transformed into a vector of 100 bits, i.e. a value of 0 or 1, corresponding to whether a certain monomer was part of the knotted arc. 

Unlike normal multi-class classification problems where the classes are mutually exclusive, here we consider a multi-label classification task, with mutually non-exclusive class labels (multiple classes per prediction)~\cite{multilabel}. To quantify the error in a multi-label classification task, we use the binary cross-entropy (BCE) function, suited to an output layer of sigmoid functions, given by
\begin{equation}
    BCE = - \frac{1}{N}\sum_{i=1}^N y_i \log(\hat{y}_i) + (1-y_i)\log(1-\hat{y}_i)
\end{equation}
where $y_i$ is the $i$\textsuperscript{th} element in the true output vector, $\textbf{y}$, $\hat{y}_i$ is the $i$\textsuperscript{th} element in the predicted output vector, \bm{$\hat{y}$}, and $N$ is the dimension of the output label, corresponding to the length of the polymer in our knot localisation task. This error is then used to optimise the model weights. 
 
Finally, to determine the accuracy of the model we converted the probabilities generated by the sigmoid function $y_{prob}$ into binary values using an Heaviside step function (${y_{pred}} = \Theta({y_{prob}}-0.5)$), and compared to the true binary value obtained using Kymoknot. The final accuracy is given by the binary accuracy, i.e.  $\text{Accuracy} = \text{correct}/\text{total}$. 

Overall, we find that the StA-trained RNNs perform extremely well, reaching above 90\% accuracy in localising any knot that we tested: the 5 simplest knot types, $0_1$, $3_1$, $4_1$, $5_1$ and $5_2$ (Fig.~\ref{fig:loc_acc}). We argue that this excellent performance relies on the effectiveness of RNNs in handling multi-scale sequential data and tracking multi-scale correlations along the polymer. This capability likely plays a major role in allowing the network to recognise that nearby monomers are more likely to be in the same knotted arc. More precisely, we find that the StA writhe representation is superior to all other descriptors, with a localisation accuracy of 93\%, confirming its potential usefulness as a tool to help in knot localisation tasks. For instance, in Fig.~\ref{fig:loc_acc}D we report the prediction and ground truth for the $4_1$ knot shown in Fig.~\ref{fig:loc_acc}A-B. In this case, the StA writhe perfectly agrees with the kymoknot ground truth, whereas the XYZ and unsigned StA writhe yield less accurate localisation predictions. 

In the SI (Fig.~S9), we also used our StA-trained RNN model to track the unknotting of a $5_1$ knot tied on an open curve. Despite the fact that the algorithm was \textit{not} trained on open curves, the results were surprisingly accurate. The model can be seen to clearly detect the presence of short knotted arcs even at the final step before complete unknotting. Once again, also in this case we find that the StA-trained model is largely superior to the XYZ-trained model.

Overall, our results highlight the power of StA and StS writhe in not only classifying but also localising knots. We acknowledge that our results are non-exhaustive and more work will be needed in the future to find the best architectures and models to optimally solve these tasks.

\section{Conclusions}

In conclusion, we have discovered that local ``segment-to-all'' and ``segment-to-segment'' writhe (Eq.~\eqref{eq:stswrithe}) are geometric descriptors of a curve that contain information about its underlying topology. Our AI-driven approach can classify, using a single quantity, complex knot topologies that would otherwise be impossible to disentangle using a single algebraic invariant. More specifically, we demonstrated, for the first time, that NNs can utilise the information encoded in StA and StS writhe to classify the curve topology significantly more accurately than what can be achieved using the Cartesian coordinates of the curve's segments or other local geometric quantities (Fig.~\ref{fig:panel0}). We hypothesise that our NNs trained on local 3D writhe representations may numerically encode a new type of geometric topological invariant. This conjecture is supported by the fact that even a simple FFNN architecture can distinguish the topology of knot mutants and composites that share several algebraic knot polynomials (Fig.~\ref{fig:panel1}). Finally, we showed that our new proposed geometric feature (Eq.~\eqref{eq:stswrithe}) is robust to more complex knots than the ones tackled in the literature so far; indeed, we have managed to classify all 250 prime knots up to 10-crossings with 95\% accuracy (Fig.~\ref{fig:panel2}). We argue that deeper NN or convolutional NN may be able to push this result further, to $>$10 crossing knots.

We stress that this method only requires a snapshot of a knot embedding with a list of 3D coordinates for each polymer segment and is trained on thermal conformations under a readily tunable temperature. For this reason, it will require longer training for longer polymers but should be essentially insensitive to the number of non-essential crossings, as shown by the excellent accuracy achieved in spherically confined polymers~\cite{Braghetto2022}. This feature is in marked contrast of standard knot topology algorithms, that take 2D projections and need to compute matrices as big as the number of crossings in a given projection, irrespectively whether essential or not~\cite{Adams1994}. 
Finally, we show that by deploying recurrent NNs, our geometric StA descriptor can also solve a knot localisation problem (Fig.~\ref{fig:loc_acc}). More work will be needed in the future to find optimal NN architectures.

We note that albeit we do not have a full understanding of how the NNs are using StA and StS writhe features to identify knots, we hypothesise that they are classifying the patterns of consecutive maxima and minima, thus capturing the entanglement of pairs of segments, accounting for their chirality and magnitude. 
This argument directly suggests that employing a distance map between segments or other geometric ``unsigned'' representations will yield lower accuracies, due to the fact that they do not capture the chiral nature of the entanglements between segments. For these reasons, we believe that StS (or StA) representations are possibly the best ones to connect the geometry a given curve embedding to its underlying topology. 
A possible limitation of this method is that it is restricted to pair-wise entanglement. Generalising the Gauss linking number to higher-order relations is itself an active field of research, and it is foreseeable that a local version of the Milnor triple linking number~\cite{Polyak1997} may be used to generate 3D tensors of Brunnian links, for example. 

In conclusion, we established that StS-trained NNs are powerful tools to accurately classify and localise knots in thermally equilibrated curves. Importantly, knot classification and localisation are achieved without any explicit calculation of Alexander or other algebraic invariants. In other words, we propose that the local writhe -- once fed through deep NNs -- yields an accurate map from the configurational space of a curve to its underlying topology.
The approach we reported in our paper naturally lends itself to be applied to protein folding~\cite{Dabrowski-Tumanski2017,Giulini2019}, DNA~\cite{Siebert2017,Goundaroulis2019} and, in general, entanglements in open curves and complex systems~\cite{Dennis2010, Everaers2004, Landuzzi2020, Stano2022, rosa2020threading, Herschberg2023, Caraglio2019}. We hope that our results will also inspire mathematicians and topologists to formulate new topological invariants based on the geometrical embedding of knotted curves. 

\section*{Acknowledgements}
DM thanks the Royal Society for support through a University Research Fellowship. This project has received funding from the European Research Council (ERC) under the European Union's Horizon 2020 research and innovation program (grant agreement No 947918, TAP). JLS thanks the Physics of Life network for a student summer bursary in 2021 providing initial funds for this research. The authors thank Enzo Orlandini, Marco Baiesi, Pawel Dabrowski-Tumanski, Ken Millett, and Luigi del Debbio for insightful discussions. The authors also acknowledge the contribution of the COST Action Eutopia, CA17139.

\bibliographystyle{apsrev4-1}
\bibliography{library} 

\end{document}